\documentclass[amsmath,amssymb,aps,letterpaper,prl,twocolumn,longbibliography,10pt]{revtex4-1}
\pdfoutput=1
\usepackage{graphicx}
\usepackage{color}
\usepackage{comment}

\usepackage{color}
\usepackage{physics}
\usepackage{xfrac}

\makeatletter \let\@keywords\@empty \let\@subject\@empty
\providecommand{\keywords}[1]{\gdef\@keywords{#1}}
\providecommand{\subject}[1]{\gdef\@subject{#1}}
\def\thetitle{\@title}
\def\theauthor{\@author}
\def\thesubject{\@subject}
\def\thedate{\@date}
\def\thekeywords{\@keywords}
\makeatother
\AtBeginDocument{%
\hypersetup{pdftitle={\thetitle}}%
\hypersetup{pdfauthor={\theauthor}}%
\hypersetup{pdfsubject={\thesubject}}%
\hypersetup{pdfkeywords={\thekeywords}}%
}

\usepackage[bookmarks=true,hyperfigures=true]{hyperref}
\usepackage{fixmath}
\usepackage{mathtools}

\providecommand{\href}[2]{#2}

\let\oldbfseries=\bfseries
\let\oldmdseries=\mdseries
\let\oldnormalfont=\normalfont
\renewcommand{\bfseries}{\oldbfseries\boldmath}
\renewcommand{\mdseries}{\oldmdseries\unboldmath}
\renewcommand{\normalfont}{\oldnormalfont\unboldmath}

\makeatletter
\newlength{\apb@width}
\newcommand{\autoparbox}[2][c]{\settowidth{\apb@width}{#2}\parbox[#1]{\apb@width}{#2}}

\makeatother

\newcommand{\bin}{\mathrm{bin}}

\begin{document}

\title{Spectrum of the Hypereclectic Spin Chain and P\'olya Counting}

\author{Changrim Ahn}
 \email{ahn@ewha.ac.kr}

\affiliation{
Department of Physics, Ewha Womans University, 
52 Ewhayeodae-gil, Seodaemun-gu, Seoul 03760, S. Korea
}

\author{Matthias Staudacher}
  \email{staudacher@physik.hu-berlin.de}

\affiliation{
Institut f\"{u}r Physik, Humboldt-Universit\"{a}t zu Berlin, \\
Zum Gro{\ss}en Windkanal 2, 12489 Berlin, Germany
}

\begin{abstract}
In earlier work we proposed a generating function that encodes the Jordan block spectrum of the integrable Hypereclectic spin chain, related to the one-loop dilatation operator of the dynamical fishnet quantum field theory. We significantly improve the expressions for these generating functions, rendering them much more explicit and elegant. In particular, we treat the case of the full spin chain without imposing any cyclicity constraints on the states, as well as the case of cyclic states. The latter involves the P\'olya enumeration theorem in conjunction with q-binomial coefficients.

\end{abstract}

\maketitle

\section{Introduction and review}

A novel integrable spin chain model was introduced in \cite{Ahn:2020zly} and called the {\it Hypereclectic spin chain}. It is a three-state model defined through an exceedingly simple nearest-neighbor non-hermitian Hamiltonian acting on a length-$L$ state space 
\begin{equation}\label{eq:hilbert3}
\underbrace{
{\mathbb C}^3\otimes {\mathbb C}^3\otimes \cdots \otimes  
{\mathbb C}^3}_{L-\mbox{\scriptsize{times}}}
\end{equation}
as
\begin{equation}\label{eq:hypereclecticH}
\mathfrak{H}=\sum_{\ell=1}^L \mathfrak{P}^{\ell,\ell+1}
\end{equation}
with periodic boundary conditions $\mathfrak{P}^{L,L+1}=\mathfrak{P}^{L,1}$, where  $\mathfrak{P}^{\ell,\ell+1}:\mathbb{C}^3\otimes \mathbb{C}^3 \rightarrow \mathbb{C}^3\otimes \mathbb{C}^3$ is a chiral permutation operator that only acts non-trivially on adjacent states $\ell,\ell+1$ as
\begin{equation}\label{eq:chiralperm}
\mathfrak{P}^{\ell,\ell+1} \ket{21}=\ket{12}\,,
\end{equation}
while killing all other eight nearest-neighbor tensor product states $\ket{12}$, $\ket{32}$, $\ket{23}$, $\ket{13}$, $\ket{31}$, $\ket{11}$, $\ket{22}$, $\ket{33}$. 
Clearly the Hamiltonian $\mathfrak{H}$ preserves the individual numbers of excitations of type 1,2 and 3.
Whereas its integrability was established in \cite{Ahn:2020zly} by proving the existence of an R-matrix and a tower of commuting higher charges, it was also demonstrated that the Bethe ansatz fails. This was shown to be closely related to the non-diagonalizability of this Hamiltonian. Instead, except for the diagonalizable two-particle subsectors, $\mathfrak{H}$ may be brought, in principle, into {\it Jordan normal form} (JNF) by a suitable similarity transform. While its generalized energy eigenvalues are all trivially zero, the resulting spectrum of Jordan blocks turned out to be extremely rich, intricate and non-trivial. Powerful combinatorial methods for determining this spectrum were introduced in \cite{Ahn:2021emp}. They rest on a compelling and extensively checked but unfortunately still unproven {\it non-shortening} conjecture. It then turned out that the spectrum for fixed excitation numbers $L, M, K$ is nicely encoded into an auxiliary partition function
\begin{equation}\label{eq:ZLMKtr}
Z^{\rm all}_{L,M,K}(q)=\Tr^{\rm all}_{L,M,K} q^{\hat{S}-\scriptstyle{\frac{1}{2}}\hat{S}_{\text{max}}}\,.
\end{equation}
Here one takes the trace of a certain {\it diagonal} counting operator $q^{\hat{S}-\scriptstyle{\frac{1}{2}}\hat{S}_{\text{max}}}$ over {\it all} states with fixed numbers of $L\!-\!M$ 1s, $M\!-\!K$ 2s and $K$ 3s. The meaning of $\hat{S}$ is as follows: Since the states 3 are non-movers w.r.t.\ the action of $\mathfrak{H}$, we may divide each spin chain configuration into $K$ {\it bins} separated by $K$ static walls (the 3s). Each bin contains a certain assortment of 2s and 1s, in some order. $\hat{S}$ then sums over the ``content'' of all $K$ bins, where the ``content'' of each bin is defined as the number of 1s (including multiple counts) to the right of the 2s within that bin. For details see \cite{Ahn:2021emp}. 
One may then decode in a computationally efficient and unique way from \eqref{eq:ZLMKtr} the sizes and multiplicities of the Jordan blocks of the sector with excitations numbers $L, M, K$. To this end, we simply expand a given \eqref{eq:ZLMKtr} into the (finite) sum
\begin{equation}\label{eq:extract}
Z^{\rm all}_{L,M,K}(q) = \sum_{j=1}^\infty N_j\,[j]_q \,,
\end{equation}
where $[j]_q$ is a $q$-number
\begin{equation}\label{eq:qnumber}
[j]_q = \frac{q^{j/2}-q^{-j/2}}{q^{1/2}-q^{-1/2}}= \sum_{k=-\frac{j-1}{2}}^{\frac{j-1}{2}} q^{k}\,.
\end{equation}
The number of Jordan blocks of length $j$ is then given by the coefficients $N_j$. Therefore, in order to determine the spectrum of the Hypereclectic spin chain, the goal is to find explicit expressions for \eqref{eq:ZLMKtr}. A consistency check should always yield for $q \rightarrow 1$ the correct number of states
\begin{equation}\label{eq:trinomial}
Z^{\rm all}_{L,M,K}(1)=\frac{L!}{(L-M)!\, (M-K)!\, K!}\,.
\end{equation}
Turning this around, we are looking for a suitable $q$-deformation of \eqref{eq:trinomial}.
Crucial first steps to compute \eqref{eq:ZLMKtr} were taken in \cite{Ahn:2021emp}. We will complete them in this letter.

In addition to preserving excitation numbers, the  Hamiltonian $\mathfrak{H}$ in \eqref{eq:hypereclecticH},\eqref{eq:chiralperm} possesses another manifest symmetry: Invariance under shifts by one lattice site. This means that we may change the unentangled tensor product basis of the state space for fixed $L, M, K$ to a new basis, where all basis vectors possess a fixed eigenvalue under the action of the shift operator. The $L$ possible shift eigenvalues are the $L$-th roots of unity $\exp(2 \pi i m/L)$, $m=0,1, \ldots ,L\!-\!1$, see e.g.\ \cite{Ahn:2021emp} for a detailed discussion. When we bring $\mathfrak{H}$ into JNF, we may take into account the shift symmetry, and find the JNF for specific sectors labelled by $m$. In this letter, we will do this for the {\it cyclic states} ($m=0$, i.e.\ the shift eigenvalue is one), which are of particular interest. Luckily, our formalism immediately carries over, c.f.\ \cite{Ahn:2021emp}, with \eqref{eq:ZLMKtr} simply replaced by
\begin{equation}\label{eq:cyclicTr}
Z^{\rm cyc}_{L,M,K}(q)=\Tr^{\rm cyc}_{L,M,K} q^{\hat{S}-\scriptstyle{\frac{1}{2}}\hat{S}_{\text{max}}}\,,
\end{equation}
where one now traces the same diagonal operator only over all cyclic states at fixed excitation numbers. The cyclic JNF is then again extracted from an expansion exactly as in \eqref{eq:extract}. However, the consistency check for $q \rightarrow 1$ now leads to
\begin{equation}\label{eq:cyclicstates}
Z^{\rm cyc}_{L,M,K}(1)=\# \left(\text{cyclic states with fixed}\,\, L, M, K\right).
\end{equation}
The counting of the number of cyclic states is significantly more sophisticated than the naive guess ``$1/L$ times the trinomial formula in \eqref{eq:trinomial}''. The correct result is given by the P\'olya enumeration theorem. It provides a generating function for these numbers, and therefore requires summing over the $L, M, K$ (we assume the presence of at least one 3), where the variables $x,y,z$ keep track of the number of 1s, 2s and 3s:
\begin{equation}
\label{eq:Polya}
\begin{aligned}
&\sum_{L=1}^\infty \sum_{M=1}^L \sum_{K=1}^M\,
x^{L-M} y^{M-K} z^K \\
&\,\,\,\,= -\sum_{n=1}^\infty \frac{\phi(n)}{n}\,\log\frac{1-x^n-y^n-z^n}{1-x^n-y^n}\,.
\end{aligned}
\end{equation}
Here $\phi(n)$ is {\it Euler's totient function}, defined as the number of positive integers less than $n$ that are coprime to $n$ (i.e.\ the number of those elements of $\{1,\ldots,n\!-\!1\}$ whose only divisor common with $n$ is 1). Clearly, if $n$ is prime, then $\phi(n)=n\!-\!1$, and if $n$ is not prime, $\phi(n)<n\!-\!1$. A very nice physicist's derivation of the general form of P\'olya's enumeration theorem is given in \cite{Spradlin:2004pp}, see also \footnote{In the notation of \cite{Spradlin:2004pp}, their generating function for the beads $z(x)$ has to be replaced by $x\!+\!y\!+\!z$ (respectively $x\!+\!y$ for the subtraction of the $K=0$ states) in order to derive \eqref{eq:Polya}, and by bin$(x,y,z,q)$ in order to derive \eqref{eq:ZcycExpansion}.}. The denominator of the logarithm's argument in \eqref{eq:Polya} serves to subtract the contributions from the $K=0$ states, i.e.\ the ones without at least one 3. Our goal is to find, in light of \eqref{eq:cyclicTr}, a suitable $q$-deformation of \eqref{eq:Polya}.

It should be pointed out that the Hypereclectic Hamiltonian $\mathfrak{H}$ is part of the one-loop dilatation operator of a strongly twisted, double-scaled deformation of $\mathcal{N}=4$ Super Yang-Mills Theory that has been christened (dynamical) fishnet theory, see \cite{Ahn:2020zly,Ahn:2021emp,Ipsen:2018fmu,Kazakov:2018gcy} for more details and references therein. In this field theory application, however, one should restrict the attention to cyclic states only, see above. We would also like to point the reader to the recent work \cite{Garcia:2021mzb,NietoGarcia:2022kqi}, where a complementary approach to determining the spectrum of the Eclectic spin chain was presented.

\section{All-sector solution of the chain}

\paragraph{The $K=1$ case}

Let us start with a single bin. This case was already solved in a completely satisfactory fashion in \cite{Ahn:2021emp}, and we will just state the result without derivation. Define
\begin{equation}\label{eq:q-binomial}
\genfrac{[}{]}{0pt}{0}{\ell+m}{m}_q:=
\prod_{k=1}^m
\frac{q^{\frac{\ell+k}{2}}-q^{-\frac{\ell+k}{2}}}{q^{\frac{k}{2}}-q^{-\frac{k}{2}}}
\end{equation}
to be the $q$-binomial coefficients in the conventions used in the theory of quantum groups. Clearly they are symmetric under the exchange $q\rightarrow q^{-1}$. For $q \rightarrow 1$ \eqref{eq:q-binomial} turns into the ordinary binomial coefficient $\binom{\ell+m}{m}$. The $K=1$ partition function in \eqref{eq:ZLMKtr} was then found to be
\begin{equation}\label{eq:K=1}
Z^{\rm all}_{L,M,1}(q)=L\, Z^{\rm cyc}_{L,M,1}(q)
\,\,\, {\rm with} \,\,\,     
Z^{\rm cyc}_{L,M,1}(q)=\genfrac{[}{]}{0pt}{0}{L-1}{M-1}_q\,.
\end{equation}
We thus know the result for both all states as well as for cyclic states. In fact, P\'olya counting is trivial for $K=1$, as the single state 3 always prevents non-trivial symmetries under the action of the shift operator. Put differently, the 3 is marking a specific site of the chain.

In light of the discussion following \eqref{eq:ZLMKtr}, we now consider the trace over all states that are composed of all distinct configurations of $1$s and $2$s with a single $3$ to the right, i.e.\ taking values in the set
\begin{equation}\label{eq:bindef}
{\mathcal A}=\{{\ket{3},\ket{13},\ket{23},\ket{113},\ket{123},\ket{213},\ket{223},\ldots}\}\,.
\end{equation}
Note that ${\mathcal A} \subset \oplus_{L=1}^\infty \left(\mathbb{C}^3\right)^{\otimes L}$.
Using \eqref{eq:cyclicTr}, \eqref{eq:K=1}, we define
\begin{equation}\label{eq:bin}
\begin{aligned}
\bin(x,y,z,q)&:=z\Tr_{\mathcal A}\left[x^{\hat\ell}y^{\hat m}
q^{\hat{S}-\scriptstyle{\frac{1}{2}}\hat{S}_{\text{max}}}
\right]
\quad \left(\hat{S}_{\text{max}}=\hat \ell\, \hat m \right)
\\
&= \sum_{L=1}^\infty \sum_{M=1}^L Z^{\rm cyc}_{L,M,1}(q)\, x^{L-M} y^{M-1} z\,,
\end{aligned}
\end{equation}
where ${\hat\ell}$, ${\hat m}$ count the numbers of $1$s resp.\ $2$s of a state 
in ${\mathcal A}$.
$\hat S$ is the ``content'' of our single bin as defined after \eqref{eq:ZLMKtr}.
We may explicitly express the function ``bin'' as a sum over a product in two different ways:
\begin{equation}\label{eq:binalt}
\begin{aligned}
\bin(x,y,z,q)
&=z \sum_{m=0}^\infty y^m \prod_{\ell=0}^m \frac{1}{1-q^{\ell-\frac{m}{2}}x}\\
&=z \sum_{\ell=0}^\infty x^\ell \prod_{m=0}^\ell \frac{1}{1-q^{m-\frac{\ell}{2}}y}\,.
\end{aligned}
\end{equation}

\paragraph{The case of general $K$}

We encode the partition functions for the Hypereclectic spin chain at fixed $L,M,K$ (all cyclicity sectors) into a grand-canonical partition function through
\begin{equation}\label{eq:ZallLMK}
Z_{\rm all}(x,y,z,q)=\sum_{L=1}^\infty \sum_{M=1}^L \sum_{K=1}^M Z^{\rm all}_{L,M,K}(q)\, x^{L-M} y^{M-K} z^K\,.
\end{equation}
Consider now an object $X$ and the generating function
\begin{equation}\label{eq:Zlog}
Z(X)=-\log(1-X)=\sum_{k=1}^\infty \frac{1}{k}\, X^k\,.
\end{equation}
Its combinatorial meaning is a sum over cyclic arrangements of $k$ objects, each with a symmetry factor of $1/k$ (or a ``necklace'' made from $k$ beads ``$X$''). We can mark one bead by the operation
\begin{equation}\label{eq:Zgeom}
X \frac{\partial}{\partial X}\, Z(X)=\frac{X}{1-X}=\sum_{k=1}^\infty X^k\,.
\end{equation}
This stands for a sum over all linearly ordered arrangements of the $k$ objects $X$ 
(i.e.\ the necklace has been opened at the marked bead, without removing the latter). In generalization of \eqref{eq:Zgeom}, we can now write down an exact expression for \eqref{eq:ZallLMK} in terms of the $K=1$ solution \eqref{eq:bin} by taking $X:={\rm bin}(x,y,z,q)$ and replacing $X \frac{\partial}{\partial X}$ by the first order differential operator ${\cal D}:=\left(x \partial_x +y \partial_y +z \partial_z \right)$:
\begin{equation}\label{eq:DZallLMK}
Z_{\rm all}(x,y,z,q)
={\cal D} \left(- \log \left(1-\bin(x,y,z,q)\right) \right). 
\end{equation}
This works because we now sum over all necklaces whose ``beads'' are replaced by ``strands'' taken from the infinite set \eqref{eq:bindef}, and where we have opened the necklace by marking a specific bead inside some strand (as opposed to an entire strand).
One easily computes from \eqref{eq:DZallLMK}
\begin{equation}\label{eq:Zall}
\begin{aligned}
Z_{\rm all}(x,y,z,q)
&=\frac{{\cal D}\, \bin(x,y,z,q)}{1-\bin(x,y,z,q)} \\
&=\sum_{k=1}^\infty \left(\bin(x,y,z,q)\right)^{k-1} \, {\cal D}\, \bin(x,y,z,q).
\end{aligned}
\end{equation}
Using and generalizing \eqref{eq:binalt}, we rewrite it as
\begin{widetext}
\begin{equation}\label{eq:ZallExpansion}
Z_{\rm all}(x,y,z,q)=
\sum_{k=0}^\infty z^{k+1}\!
\left(\sum_{m=0}^\infty y^m \prod_{\ell=0}^m \frac{1}{1-q^{\ell-\frac{m}{2}}x}\right)^k\!
\left(\sum_{m=0}^\infty (m\!+\!\tfrac{1}{2})\, y^m \prod_{\ell=0}^m \frac{1}{1-q^{\ell-\frac{m}{2}}x}+
\sum_{\ell=0}^\infty (\ell\!+\!\tfrac{1}{2})\, x^\ell \prod_{m=0}^\ell \frac{1}{1-q^{m-\frac{\ell}{2}}y}\right).
\end{equation}
\end{widetext}
This explicit grand-canonical generating function vastly improves eq.(4.30) in \cite{Ahn:2021emp}: We no longer need to sum over partitions, nor consider any implicit symmetry factors.

\section{Cyclic sector solution of the chain}

Our derivation of the partition function of the full state space, using a ``second type of spin chain'' of length $K$ (instead of $L$), and whose ``spins = bins=strands'' take values in the (infinite) list of $\mathcal A$ in \eqref{eq:bindef} (instead of the set $\{1,2,3\}$) may also be nicely adapted to to the cyclic state space. We simply apply the P\'olya enumeration theorem to this ``non-compact'' spin chain, with infinitely many possible states per site. In fact, as also explained in the transparent derivation in \cite{Spradlin:2004pp}, the theorem is also valid for this case: What enters is the generating function (see again \cite{Note1}) of the states at one site: In the second, non-compact chain of length $K$ this is essentially $\bin(x,y,z,q)$, instead of the generating function $x\!+\!y\!+\!z$ of the first, compact chain of length~$L$. P\'olya's theorem then yields
\begin{equation}
\label{eq:ZcycExpansion}
\begin{aligned}
&Z_{\rm cyc}(x,y,z,q)= -\sum_{n=1}^\infty \frac{\phi(n)}{n}\,\log\left(1-\bin(x^n,y^n,z^n,q^n)\right)\\
&=\sum_{L=1}^\infty \sum_{M=1}^L \sum_{K=1}^M  Z^{\rm cyc}_{L,M,K}(q)\, x^{L-M} y^{M-K} z^K.
\end{aligned}
\end{equation}
One may check, that this expression indeed reduces back to \eqref{eq:Polya} for $q \rightarrow 1$. 
In fact, \eqref{eq:ZcycExpansion} is the $q$-deformation of \eqref{eq:Polya} we had been looking for. As in \eqref{eq:extract}, this determines the Jordan spectrum of the cyclic sector after expanding $Z^{\rm cyc}_{L,M,K}(q)$ into $q$-numbers.

\section{An Example}
 
Let us illustrate the power of the above generating functions in one concrete example: 
$L\!=\!9, M\!=\!6, K\!=\!3$.
It corresponds to the Hypereclectic spin chain of length nine, with three 1s, 2s and 3s each.

\paragraph{All sectors} Using Mathematica\texttrademark, one easily finds within seconds the coefficient of $x^3 y^3 z^3$ in the series expansion of the explicit formula \eqref{eq:ZallExpansion}, i.e.\ in light of \eqref{eq:ZallLMK}
\begin{equation}
\label{eq:Zall963}
\begin{aligned}
&Z^{\rm all}_{9,6,3}(q)= 9 q^{-9/2}\!+\!9 q^{-7/2}\!+\!36 q^{-3}\!+\!36 q^{-5/2}\!+\!72 q^{-2} \\
&\!+\!156 q^{- 3/2}\!+\!162 q^{-1}\!+\!234 q^{-1/2}\!+\!252\!+\!234 q^{1/2}\!+\!162 q \\
&\!+\!156 q^{3/2}\!+\!72 q^2\!+\!36 q^{5/2}\!+\!36 q^3\!+\!9 q^{7/2}\!+\!9 q^{9/2}\,.
\end{aligned}
\end{equation}
From \eqref{eq:trinomial} there are $Z^{\rm all}_{9,6,3}(1)=9!/(3!)^3=1680$ states in this sector.
We immediately rewrite this with \eqref{eq:extract}, \eqref{eq:qnumber}  as
\begin{equation}
\label{eq:extract963}
\begin{aligned}
Z^{\rm all}_{9,6,3}(q)=& 90\, [1]_q + 78\, [2]_q+90\, [3]_q + 120\, [4]_q + 36\, [5]_q \\
&+ 27\, [6]_q + 36\, [7]_q + 9\, [10]_q\,.
\end{aligned}
\end{equation}
From this we can read of the JNF:
\begin{equation}
\label{eq:JNF963}
{\rm JNF}^{\rm all}_{9,6,3}=(10^9, 7^{36}, 6^{27}, 5^{36}, 4^{120}, 3^{90}, 2^{78}, 1^{90}).
\end{equation}
The notation is the same as in \cite{Ahn:2021emp}, i.e.\ there are 9 Jordan blocks of size 10, 36 blocks of size 7, and so on. We have verified this JNF with a Mathematica\texttrademark\ program directly applying the linear algebra method detailed in eqs.\ (4.24)-(4.26) of \cite{Ahn:2020zly} to the Hamiltonian $\mathfrak{H}$ in \eqref{eq:hypereclecticH}.

\paragraph{Cyclic sector}

Using once again Mathematica\texttrademark, one quickly finds the coefficient of $x^3 y^3 z^3$ in 
\eqref{eq:ZcycExpansion}:
\begin{equation}
\label{eq:Zcyc963}
\begin{aligned}
&Z^{\rm cyc}_{9,6,3}(q)= q^{-9/2}\!+\!q^{-7/2}\!+\!4 q^{-3}\!+\!4 q^{-5/2}\!+\!8 q^{-2} \\
&\!+\!18 q^{- 3/2}\!+\!18 q^{-1}\!+\!26 q^{-1/2}\!+\!28\!+\!26 q^{1/2}\!+\!18 q \\
&\!+\!18 q^{3/2}\!+\!8 q^2\!+\!4 q^{5/2}\!+\!4 q^3\!+\!q^{7/2}\!+\!q^{9/2}\,.
\end{aligned}
\end{equation}
For $q\!\!=\!\!1$ this agrees with the number of $Z^{\rm cyc}_{9,6,3}(1)=188$ cyclic states with with three 1s, 2s and 3s each, as predicted by the ``undeformed'' P\'olya counting formula \eqref{eq:Polya}. Note that (see paragraph above) $1680/9\neq 188$: P\'olya counting is non-trivial in this sector. In consequence, we also must have $Z^{\rm all}_{9,6,3}(q) \neq 9 \times Z^{\rm cyc}_{9,6,3}(q)$. In fact, we may reexpress \eqref{eq:Zcyc963} as
\begin{equation}
\label{eq:extract963cyc}
\begin{aligned}
Z^{\rm cyc}_{9,6,3}(q)=& 10\, [1]_q + 8\, [2]_q+10\, [3]_q + 14\, [4]_q + 4\, [5]_q \\
&+ 3\, [6]_q + 4\, [7]_q + [10]_q\,,
\end{aligned}
\end{equation}
which should be compared to \eqref{eq:extract963}. The JNF then reads
\begin{equation}
\label{eq:JNF963cyc}
{\rm JNF}^{\rm cyc}_{9,6,3}=(10, 7^4, 6^3, 5^4, 4^{14}, 3^{10}, 2^8, 1^{10}).
\end{equation}
It should be compared to \eqref{eq:JNF963}.

\section{Conclusion and open problems}

In conclusion, we have exactly computed the partition functions
$Z^{\rm all}_{L,M,K}(q)$ \eqref{eq:ZLMKtr} and $Z^{\rm cyc}_{L,M,K}(q)$ \eqref{eq:cyclicTr} of the Hypereclectic spin chain model. This was done by providing explicit formulas for their grand canonical versions $Z_{\rm all}(x,y,z,q)$\eqref{eq:ZallLMK} and $Z_{\rm cyc}(x,y,z,q)$ \eqref{eq:ZcycExpansion}. 
The main advantage over the formula (4.30) for {\it all} states in \cite{Ahn:2021emp} is that we no longer need 
the implicit symmetry factors $S_{\boldsymbol{\ell},\boldsymbol{m}}$. 
Also, we can now treat the cyclic sector in generality.

Given the arguments in \cite{Ahn:2021emp}, our expressions should then provide the exact spectrum of Jordan blocks of the Hypereclectic spin chain. However, this still hinges on our {\it non-shortening conjecture} in \cite{Ahn:2021emp}.
While we have extensively checked this conjecture --- and the 
$L\!=\!9, M\!=\!6, K\!=\!3$ example worked out for this letter is another non-trivial test --- it would still be important to {\it prove} it.
Particularly desirable would be a proof based on the {\it integrability} of the Hypereclectic spin chain \cite{Ahn:2020zly}. Here we find the appearance of $q$-numbers \eqref{eq:qnumber} and $q$-binomials \eqref{eq:q-binomial}, first noted in \cite{Ahn:2021emp}, very promising, as they are ubiquitous in the theory of quantum groups. The latter form the mathematical foundation of many integrable models.

The Hypereclectic Hamiltonian $\mathfrak{H}$ in \eqref{eq:hypereclecticH} is a special case of a more general integrable, non-hermitian three-state model, the {\it Eclectic spin chain} of \cite{Ahn:2020zly}. In its cyclic sector, it is related to the one-loop dilatation operator of ``dynamical fishnet theory'', a strongly twisted, double-scaled a three-parameter deformation of $\mathcal{N}=4$ Super Yang-Mills Theory, see \cite{Ahn:2020zly,Ahn:2021emp,Ipsen:2018fmu,Kazakov:2018gcy} for details and references. Its Hamiltonian 
reads
\begin{equation}\label{eq:Hec}
\mathcal{H}=\sum_{\ell=1}^L \left(\xi_1\, \mathfrak{P}^{\ell,\ell+1}_1 + \xi_2\, \mathfrak{P}^{\ell,\ell+1}_2 + \xi_3\, \mathfrak{P}^{\ell,\ell+1}_3 \right) .
\end{equation}
Dropping the adjacent-site labels $\ell,\ell+1$, the (now three) chiral permutation operators act as
\begin{equation}
\mathfrak{P}_1\ket{32}=\ket{23},\quad \mathfrak{P}_2 \ket{13}=\ket{31}, \quad \mathfrak{P}_3\ket{21}=\ket{12}\,.
\end{equation}
They anihilate all other nearest-neighbor tensor product states, respectively. We recover \eqref{eq:hypereclecticH}, \eqref{eq:chiralperm} for $\xi_1=\xi_2=0$, $\xi_3=1$. In \cite{Ahn:2020zly,Ahn:2021emp} a rather surprising {\it universality hypothesis} was formulated: If the filling conditions $K \leq M\!-\!K \leq L\!-\!M$ are satisfied, the spectrum of Jordan blocks of the eclectic $\mathcal{H}$ in \eqref{eq:Hec} at ``generic'' values of the parameters $\xi_j$ is identical to the spectrum of the hypereclectic $\mathfrak{H}$ in \eqref{eq:hypereclecticH}. (If the filling conditions do not hold, we may always satisfy them after a suitable synchronous permutation of the three state- and parameter labels.) If this hypothesis holds, along with the above non-shortening conjecture, then we have also found in this letter the full solution of Eclectic spin chain's spectral problem for generic (i.e.\ not fine-tuned) $\xi_j$. Once again, we hope that integrability might provide the means to prove this.

Finally, it would be interesting to interpret our results within the framework of dynamical fishnet theory \cite{Kazakov:2018gcy}, believed to be a logarithmic conformal field theory (see \cite{Ahn:2021emp} for further information). For example, is the spectrum of Jordan blocks we found preserved beyond one loop?


\begin{acknowledgments}
\paragraph{Acknowledgements}
%
We thank
Luke Corcoran
for inspiring discussions, Moritz Kade 
for his excellent Mathematica program on the JNF of the Hypereclectic spin chain, and both of them for useful comments on the draft.
MS thanks Ewha Womans University for hospitality. This project received funding from NRF grant (NRF- 2016R1D1A1B02007258) (CA, MS)
\end{acknowledgments}

\bibliography{C_M_Polya}

\end{document}